\let\includefigures=\iffalse
%
\let\useblackboard=\iftrue
%
%
\newfam\black
\input harvmac
\includefigures
\message{If you do not have epsf.tex (to include figures),}
\message{change the option at the top of the tex file.}
\input epsf
\def\figin{\epsfcheck\figin}\def\figins{\epsfcheck\figins}
\def\epsfcheck{\ifx\epsfbox\UnDeFiNeD
\message{(NO epsf.tex, FIGURES WILL BE IGNORED)}
\gdef\figin##1{\vskip2in}\gdef\figins##1{\hskip.5in}
\else\message{(FIGURES WILL BE INCLUDED)}%
\gdef\figin##1{##1}\gdef\figins##1{##1}\fi}
\def\DefWarn#1{}
\def\figinsert{\goodbreak\midinsert}
\def\ifig#1#2#3{\DefWarn#1\xdef#1{fig.~\the\figno}
\writedef{#1\leftbracket fig.\noexpand~\the\figno}%
\figinsert\figin{\centerline{#3}}\medskip\centerline{\vbox{\baselineskip12pt
\advance\hsize by -1truein\noindent\footnotefont{\bf Fig.~\the\figno:} #2}}
\bigskip\endinsert\global\advance\figno by1}
\else
\def\ifig#1#2#3{\xdef#1{fig.~\the\figno}
\writedef{#1\leftbracket fig.\noexpand~\the\figno}%
\global\advance\figno by1}
\fi
\useblackboard
\message{If you do not have msbm (blackboard bold) fonts,}
\message{change the option at the top of the tex file.}
\font\blackboard=msbm10 scaled \magstep1
\font\blackboards=msbm7
\font\blackboardss=msbm5
\textfont\black=\blackboard
\scriptfont\black=\blackboards
\scriptscriptfont\black=\blackboardss

\else

\fi
%
\def\yboxit#1#2{\vbox{\hrule height #1 \hbox{\vrule width #1
\vbox{#2}\vrule width #1 }\hrule height #1 }}
\def\fillbox#1{\hbox to #1{\vbox to #1{\vfil}\hfil}}
\def\ybox{{\lower 1.3pt \yboxit{0.4pt}{\fillbox{8pt}}\hskip-0.2pt}}

\def\comments#1{}

\def\eps{\epsilon}

\def\Tr{{{\rm Tr~ }}}

\def\tr{{\rm Tr\ }}
\def\det{{\rm det\ }}

\def\ket#1{|#1\rangle}
\def\vev#1{\langle{#1}\rangle}

\def\CN{{\cal N}}

\def\II{\relax{I\kern-.10em I}}
\def\IIa{{\II}a}
\def\IIb{{\II}b}

\def\IZ{\relax\ifmmode\mathchoice
{\hbox{\cmss Z\kern-.4em Z}}{\hbox{\cmss Z\kern-.4em Z}}
{\lower.9pt\hbox{\cmsss Z\kern-.4em Z}}
{\lower1.2pt\hbox{\cmsss Z\kern-.4em Z}}\else{\cmss Z\kern-.4em
Z}\fi}
\def\IB{\relax{\rm I\kern-.18em B}}
\def\IC{{\relax\hbox{$\inbar\kern-.3em{\rm C}$}}}
\def\ID{\relax{\rm I\kern-.18em D}}
\def\IE{\relax{\rm I\kern-.18em E}}
\def\IF{\relax{\rm I\kern-.18em F}}
\def\IG{\relax\hbox{$\inbar\kern-.3em{\rm G}$}}
\def\IGa{\relax\hbox{${\rm I}\kern-.18em\Gamma$}}
\def\IH{\relax{\rm I\kern-.18em H}}
\def\II{\relax{\rm I\kern-.18em I}}
\def\IK{\relax{\rm I\kern-.18em K}}
\def\IP{\relax{\rm I\kern-.18em P}}

%

\def\inbar{\,\vrule height1.5ex width.4pt depth0pt}
\def\mod{{\rm mod}}

\font\cmss=cmss10 \font\cmsss=cmss10 at 7pt
\def\IR{\relax{\rm I\kern-.18em R}}

\def\BR{\IR}
\def\BZ{\IZ}

\def\BR{\IR}
\def\BC{\IC}

\def\lp10{l_P^{10}}
\def\lp11{l_P^{11}}
\def\R11{R_{11}}

\Title{\vbox{\baselineskip12pt\hbox{hep-th/9807235}
\hbox{RU-98-35}}}
{\vbox{
\centerline{D-branes and Discrete Torsion} }}
\centerline{Michael R. Douglas}
\medskip
\centerline{Department of Physics and Astronomy}
\centerline{Rutgers University }
\centerline{Piscataway, NJ 08855--0849}
\medskip\centerline{\it and}
\medskip\centerline{Institut des Hautes \'Etudes Scientifiques}
\centerline{Le Bois-Marie, Bures-sur-Yvette, 91440 France}
\centerline{\tt mrd@physics.rutgers.edu}
\bigskip
\noindent
We show that discrete torsion is implemented in a D-brane world-volume
theory by using a projective representation of the orbifold point group.
We study the example of $\BC^3/\BZ_2\times \BZ_2$ and show that
the resolution of singularities agrees with that proposed by Vafa
and Witten.
A new type of fractional brane appears.

\Date{July 1998}
%
\lref\dm{M. R. Douglas and G. Moore, hep-th/9603167;
J. Polchinski, hep-th/9606165;
C. Johnson and R. Myers, hep-th/9610140}
\lref\gp{E. Gimon and J. Polchinski, Phys. Rev. D54 (1996) 1667-1676,
hep-th/9601038.}
\lref\dgm{M. R. Douglas, B. Greene and D. R. Morrison,
Nucl. Phys. {\bf B506} (1997) 84, hep-th/9704151;
M. R. Douglas and B. R. Greene, hep-th/9707214;
T. Muto, hep-th/9711090}
\lref\ztwo{B. R. Greene, hep-th/9711124;
S. Mukhopadhyay and K. Ray, hep-th/9711131.}
\lref\mohri{K. Mohri, hep-th/9707012, hep-th/9806052.}
\lref\bfss{T. Banks, W. Fischler, S. H. Shenker and L. Susskind,
Phys.Rev. {\bf D55} (1997) 6382, hep-th/9610043.}
\lref\dkps{M. R. Douglas, D. Kabat, P. Pouliot and S. Shenker,
Nucl.Phys. {\bf B485} (1997) 85, hep-th/9608024.}
\lref\egs{M. R. Douglas, hep-th/9612126;
D.-E. Diaconescu and J. Gomis, hep-th/9707019;
D.-E. Diaconescu, M. R. Douglas and J. Gomis, hep-th/9712230;
M. R. Douglas, H. Ooguri and S. H. Shenker, hep-th/9702203;
M. R. Douglas and H. Ooguri, hep-th/9710178.}
\lref\kron{P. Kronheimer, J. Diff. Geom. {\bf 28}1989)665.}
\lref\Ab{P.S. Aspinwall, TASI 96, hep-th/9611137.}
\lref\witten{E. Witten, Nucl. Phys. B443 (1995) 85; hep-th/9503124.}
\lref\hm{J. Harvey and G. Moore, hep-th/9609017.}
\lref\ks{S. Kachru and E. Silverstein, hep-th/9802183;
A. Lawrence, N. Nekrasov, C. Vafa, hep-th/9803015.}
\lref\bkv{M. Bershadsky, Z. Kakushadze and C. Vafa, hep-th/9803076;
M. Bershadsky and Z. Kakushadze, hep-th/9803249.}
\lref\vafa{C. Vafa, Nucl. Phys. B273 (1986) 592.}
\lref\vw{C. Vafa and E. Witten, 
J. Geom. Phys. 15 (1995) 189--214, hep-th/9409188.}
\lref\cds{A. Connes, M. R. Douglas and A. Schwarz, hep-th/9711162.}
\lref\dh{M. R. Douglas and C. Hull, hep-th/9711165.}
\lref\hw{P.-M. Ho and Y.-S. Wu, hep-th/9801147.}
\lref\karp{G. Karpilovsky, {\it Projective representations of finite
                     groups} M. Dekker, 1985;
{\it The Schur multiplier}, Clarendon Press, 1987.}
\lref\inprog{M. R. Douglas and A. Morosov, work in progress.}

\newsec{Introduction}

D-branes on resolved orbifolds provide a simple
example of how geometry at short distances
arises as the low energy configuration space
of a world-volume gauge theory \refs{\dm,\dgm,\ztwo,\mohri,\dkps},
and provide a starting point for the definition
of Matrix theory on these spaces \refs{\bfss,\egs}.
They are described by gauge theories which are projections
of maximally supersymmetric gauge theory and share some of the
quantum properties of this theory; for example in the large $N$ limit
they are fixed point theories \refs{\ks,\bkv}.

Another variation on the orbifold story is the possiblity of discrete
torsion \vafa\ and in this note we derive the corresponding D-brane
world-volume gauge theories.  The basic construction is to embed
a projective representation of the point group in the gauge group.
This possibility arose in the earliest work \refs{\gp,\dm} and more
recently was revived by Ho and Wu \hw\ %
as a generalization of Matrix theory on orbifolds,
by analogy with the new toroidal compactifications of Matrix theory in \cds.

In fact it is easy to show that the definition of discrete torsion in
\vafa\ immediately leads to this prescription (this is
very analogous to \dh).
After giving this argument, we work out the case of 
$\BC^3/\BZ_2\times\BZ_2$ in detail, reproducing the picture of \vw.
The analysis is easy to generalize and we do $\BC^3/\BZ_n\times\BZ_n$
and some other cases in \inprog.

\newsec{Discrete torsion and projective representations}

An orbifold $M/\Gamma$ in string theory is defined by quantizing
string theory on a manifold $M$ with $\Gamma$ symmetry, 
adding twisted sectors with strings closed up to the action of $\Gamma$,
and then projecting the single string Hilbert space
to the $\Gamma$-invariant subspace.
In terms of the world-sheet functional integral, we 
consider maps from a genus $g$ world-sheet $\Sigma$ to $M$,
in which each homology class of one-cycle $c$ of $\Sigma$ can be twisted
by a group element $g(c)$, and sum over all such sets of twists.
(We consider only abelian $\Gamma$ here).

D-branes can be added to this picture by starting with ``image'' branes
on $M$ and taking the action of $\Gamma$ to permute the images in an
appropriate way.  The only ``twisted open string'' sectors needed for
consistency are the strings stretched from a brane to its images.
These combine into a larger rank gauge theory to which we apply the
orbifold projection.

In generality, we can take a configuration of $N$ coincident branes on $M$ and
postulate a space-time action $r(g)$ and
a representation $\gamma(g)$ of $\Gamma$ in the gauge group.
For $M=\BR^k$, the world-volume theory is maximally supersymmetric
Yang-Mills theory with fields $\phi \equiv (A_\mu(x), X^i(x) \chi(x))$
projected as
\eqn\proj{\gamma^{-1}(g) \phi \gamma(g) = r(g) \phi. }

Discrete torsion is associated with a two-cocycle,
an element of $H^2(\Gamma,U(1))$.  Let us denote it $\eps(g,h)$.
The simplest non-trivial example is $\Gamma=\BZ_n\times \BZ_n$ for which
$H^2(\Gamma,U(1))\cong \BZ_n$.  An explicit generator of this group is
\eqn\cocycle{
\eps_1((p,q),(p',q')) = \zeta^{(pq'-p'q)}
}
where $\zeta=e^{\pi i/n}$ ($n$ even)\footnote*{
To avoid confusion, note that this differs from the expression in \vw,
where (for example) $\zeta=-1$ for $n=2$.
This is a trivial cocycle (one which can be absorbed into a 
redefinition $g\rightarrow e^{i\theta(g)} g$).
As mentioned in a footnote in \vafa\ and as
we will see below, the weights applied to the
closed string partition function (as used in \vw) are
given by the square of a two-cocycle.} 
or $\zeta=e^{2\pi i/n}$ ($n$ odd).
The general element can be obtained as 
$\eps_k(g,h)=(\eps_1(g,h))^k$.

Now there is a natural place to insert a two-cocycle in \proj:
we simply make $\gamma(g)$ a projective representation, satisfying
\eqn\projrep{
\gamma(g)\gamma(h) = \eps(g,h) \gamma(gh).
}
Since we are only using the adjoint action of $\gamma$ in \proj,
this is compatible with the group law, and we conclude that each
projective representation leads to a sensible quotient gauge theory.

Any projective representation determines a two-cocycle and in this
sense we are incorporating enough information to describe
discrete torsion; on the other hand there are many
projective representations associated to the same two-cocycle
so we cannot say that this identification is beyond question.
We will come back to the interpretation of this additional choice later.

We now argue that this prescription is not arbitrary but is
in fact required by consistency with the original definition of
orbifold with discrete torsion.
Although a general argument can be made, its structure is clear
from considering a particular two-loop nonplanar
gauge theory diagram $\Sigma$.
This is a diagram formed by tying two three-point vertices $\Tr ABC$
and $\Tr CBA$ together with the propagators $\vev{AA}$,$\vev{BB}$ and
$\vev{CC}$.  Equivalently one takes the three legs of a vertex $\tr ABC$
and twists each of them, $A\rightarrow A^{tr}$ etc., before attaching
to another vertex $\tr ABC$.
This comes from a string world-sheet
which is a genus one surface with a single boundary.
The cycles on this surface can be built by attaching propagators
$AA$, $BB$ and $CC$ at vertices.
A cycle which passes through two propagators
is not a boundary but instead
is a combination of $a$ and $b$ cycles; two cycles which pass through\
a single common propagator intersect once (because of the twist).

We can implement the projection \proj\ in gauge theory
by summing over group elements acting on the propagators.
This leads to the amplitude 
\eqn\amp{\sum_{g_1g_2g_3} 
\tr \gamma(g_1)\gamma^{-1}(g_2)\gamma(g_3)
\gamma^{-1}(g_1)\gamma(g_2)\gamma^{-1}(g_3)
\vev{r(g_1)A~r(g_2)B~ r(g_3)C} ~
\vev{ABC}.
}
By using the symmetry of the theory under the group action
we can set one of the group elements to $1$, say $g_3=1$.
The trace can then be evaluated using \projrep: it is
$$\eps(g_1,g_2^{-1})\eps(g_1^{-1},g_2) .
$$
From the definition \cocycle\ we have
$\eps(g_1^{-1},g_2) = \eps(g_1,g_2^{-1})$
and we see that the diagram is weighed by the square of the two-cocycle.
In fact $g_1$ and $g_2$ are the twists about the cycles $AACC$ and
$BBCC$ with intersection number $1$, i.e. $a$ and $b$ cycles.

Thus the projectivity of the group representation translates directly
into the phase factor of discrete torsion.  It is easy to generalize
this argument to certain higher genus surfaces, 
for example to the $k$-loop ``ladder'' diagram with
$k+1$ twists which corresponds to a genus $k/2$ world-sheet with
one ($k$ even) or two ($k$ odd) boundaries.
It would be interesting to complete the argument for general
diagrams and for the nonabelian case.

\newsec{The $\BZ_2\times \BZ_2$ example}

Let the generators
of $\BZ_2\times\BZ_2$ be $g$ and $h$ acting on the complex coordinates
$z^i$, $1\le i\le 3$ as
\eqn\gaction{\eqalign{
g:(z^1,z^2,z^3)\rightarrow (z^1,-z^2,-z^3) \cr
h:(z^1,z^2,z^3)\rightarrow (-z^1,z^2,-z^3) .
}}
Acting on $\BC^3$, an individual group element leaves a fixed line
(in algebraic geometer's language -- in real terms, a two-dimensional plane)
of $\BC^2/\BZ_2$ singularities.  These fixed lines intersect in
a point singularity at the origin.
The orbifold can be parameterized by the $\Gamma$-invariant variables
$$m_i \equiv (z^i)^2; \qquad  b \equiv z^1z^2z^3 $$
satisfying 
\eqn\invvars{m_1m_2m_3 = b^2.}

The theory without discrete torsion was studied in \ztwo;
it is a $U(N)^4$ gauge theory with bifundamental matter in the 
$N_i\times\bar N_j$ for $1\le i\ne j\le 4$ for all $i,j$, admitting
Fayet-Iliopoulos terms which completely resolve the singularity.

There is a single closed string theory with discrete torsion.
The non-trivial two-cocycle is $\eps(g,h)=-\eps(h,g)=i$.
A D-brane theory on this orbifold will be determined by a choice of 
projective representation $\gamma$ realizing this cocycle.

In fact there is a unique such irreducible representation \karp.  
It is two dimensional with
\eqn\reptwo{
\gamma(1) = 1; \qquad \gamma(g) = \sigma_1; \qquad 
\gamma(h) = \sigma_2; \qquad \gamma(gh) = \sigma_3
}
where $\sigma_i$ are the usual Pauli matrices.

The general representation is the tensor product $\gamma \otimes M_N$;
in other words $N\times N$ matrices with matrix elements taken from
\reptwo.
For example, the regular representation is defined by the
matrix elements $\gamma_r(g)\ket{h} = \eps(g,h)\ket{gh}$;
it is unitarily equivalent to $\gamma \otimes M_2$.

Thus the gauge theories describing D-branes near this singularity
are the theory
obtained by solving \proj\ with the representation \reptwo, and
matrix versions of this theory.  The solution of \proj\ is
\eqn\soltwo{ A \propto 1; \qquad Z^i \propto \sigma_i. }
The matrix version of this theory is obtained by tensoring these
with $N\times N$ matrices satisfying the standard reality conditions,
and substituting these solutions into the $U(2N)$ SYM Lagrangian.

The final result (in $\CN=1$ superfield notation)
is a $U(N)$ gauge theory with three chiral superfields $Z^1$, $Z^2$
and $Z^3$ in the $N\times \bar N$.  There is a superpotential
\eqn\superp{W = \Tr Z^1Z^2Z^3 + Z^2Z^1Z^3 .}
The F-flatness conditions are
\eqn\fflat{
Z^i Z^j + Z^j Z^i = 0 \qquad
\forall i\ne j.
}

We first consider the moduli space for $N=2$.
Since this is the regular representation 
we expect it to be $M/\Gamma$.
In this case it is useful to decompose the chiral superfields
into the singlets
$s^i = \Tr Z^i$ and a traceless $Z^i$, in terms of which the F-flatness
conditions become ${1\over 4}s^is^j + \Tr Z^iZ^j = 0$ and
$s^i Z^j + s^j Z^i = 0$.
The only solutions of these have $s_i=0$.
A solution of these and the D-flatness conditions
$\sum_i [Z^i,Z^{i+}]=0$ is
\eqn\soltwo{Z^i = z^i \tau^i }
where $\tau^i$ are another copy of the Pauli matrices, acting in the
gauge $U(2)$ representation space.
This vacuum leaves as unbroken discrete gauge symmetries the $SU(2)$
group elements $i\tau_i$ which act on the coordinates $z^i$ as \gaction,
so these solutions reproduce the expected moduli space.

To prove that these exhaust the gauge-equivalence classes of solutions, 
we consider the moduli space as parameterized by the gauge invariant
polynomials in the (traceless) $Z^i$.  These are
$$M_{ij} \equiv \Tr Z^i Z^j$$
and\
$$B \equiv \Tr Z^1[Z^2,Z^3]$$
satisfying the relation
$$\det M = B^2.$$
This gives a six complex dimensional space  which
the conditions \fflat\ reduce to the three complex dimensional variety
$$M_{11} M_{22} M_{33} = B^2.$$
The natural gauge-invariant coordinates on the moduli space are
also the natural $\Gamma$-invariant coordinates, a fact which will
be useful when we study the parameters which resolve the singularity.

We now come back to $N=1$.  The moduli space has three branches:
$z^1=z^2=0$,
$z^2=z^3=0$ and $z^1=z^3=0$, in other words the three fixed planes.
Now the fixed planes are really $\BC \times \BC^2/\BZ_2$ so
as in \egs\ this could describe some sort of object wrapped about
a hidden cycle.  We will return to discuss it after we
understand the resolution, in the next section.

For $N>2$ we can clearly take a direct sum of solutions \soltwo\ to
describe $N/2$ points in $M/\Gamma$.  That these are the only solutions
we believe follows from the fact that there is a unique irreducible
representation of the Clifford algebra \fflat, but would be worth a proof.

For $p$-branes with $p<3$, the gauge theory will also have a Coulomb
branch describing $N$ copies of the $N=1$ object, as well as mixed
branches.

\newsec{Resolving the singularity.}

In string theory, turning on twisted closed string moduli
in general has the effect of resolving orbifold singularities.
On the other hand, Vafa and Witten studied this example \vw\  and concluded
that, since the twisted closed string modes for $T^6/\BZ_2\times \BZ_2$
are in one-to-one correspondance with {\it planes} fixed by a single
group element, their likely interpretation is to resolve these fixed
plane singularities, leaving the fixed points unresolved.

We proceed to verify that this is what is seen by a D-brane.
The main observation we need is that the twist sector moduli enter
with the world-sheet quantum numbers of complex structure deformations,
i.e. elements of $H^{2,1}(M)$.  This is because the discrete torsion
modifies the orbifold projection, as explained clearly in \vw.
They are therefore complex variables
which can naturally couple to the superpotential.  Let us call these
moduli $\zeta_i$ where $i$ labels the twist in the notation given in
\reptwo.  A natural gauge invariant coupling to the D-brane theory is
then
$$\Delta W = \sum_i \zeta_i \Tr Z^i$$
which couples the $i$'th twisted sector to the leading operator
fixed under the $i$'th group element.
From world-sheet considerations as in \dm, one can see that
such a coupling is possible
(we have not explicitly verified that it is non-zero).

This coupling deforms the F-flatness conditions to
$$Z^i Z^j + Z^j Z^i = \epsilon^{ijk} \zeta_k .$$
On the other hand, since there are no $U(1)$ gauge fields associated with
twist sectors, the D-flatness conditions cannot be deformed by these
parameters.

Let us consider the $N=2$ moduli space.  One can check that
$\Tr Z^i=0$ still and in terms of the $SU(2)$ invariants $M^{ij}$
and $B$ the moduli space is
\eqn\newvar{\eqalign{
M^{ij} &=  \epsilon^{ijk} \zeta_k; \qquad i\ne j \cr
\det M &= B^2 .
}}
This is completely compatible with the resolution suggested in \vw.
Let us consider for example $\zeta_1=\zeta_2=0$ and $\zeta_3\ne 0$.
Then we have
$$(M_{11} M_{22}-\zeta_3^2) M_{33} = B^2$$
which is a deformation of the singularity first order in $M_{33}$,
resolving the fixed plane at $Z^1=Z^2=0$ but retaining a singularity
at $M_{11}M_{22}=\zeta_3^2$, $M_{33}=B=0$.  The other two fixed
planes have deformed into a single smooth fixed plane.

A complete resolution (all $\zeta\ne 0$) will move the singularity
to 
\eqn\singloc{
M_{ii}=\zeta_j\zeta_k/\zeta_i \qquad (j\ne k\ne i).  
}
Expanding \newvar\ about this point, $M_{ii}=x_i +\zeta_j\zeta_k/\zeta_i$,
leads to
$$B^2 = \sum_{i<j} {\zeta_i\zeta_j\over\zeta_k} x_i x_j + O(x^3) .
$$
The quadratic form appearing here has determinant $\zeta_1\zeta_2\zeta_3$
and we conclude that the remaining 
singularity is a conifold singularity whenever
this is nonzero.

The $N=1$ deformed moduli space changes its character under these
deformations.  If $\zeta_1\zeta_2\zeta_3=0$ but some $\zeta\ne 0$,
we have (say) $z^1z^2=\zeta_3$ and $z^3=0$ so the object is confined
to the remaining fixed plane.  
If $\zeta_1\zeta_2\zeta_3\ne 0$ the moduli space degenerates to the point
\singloc.  

The $N\ge 2$ Coulomb branch (for $p<3$)
can also be interpreted as describing such
objects (``fractional branes''), in
the same way that the Coulomb branch for conventional
orbifold theories was interpreted in \egs.
As in that case, breaking the $U(N)$ gauge symmetry to $U(1)^N$ leaves
the right fermion zero modes to describe $N$ BPS objects.
Also as in that case, two of the objects can annihilate (the transition from
Coulomb to Higgs branch) to produce a D$p$-brane.
Unlike that case, there is no conserved quantum number 
corresponding to wrapping number (except for $N \mod 2$ which can be
regarded as a $\BZ_2$ quantum number).
This is consistent with the fact that 
there is no additional gauge field in the supergravity sector for them
to couple to.
Furthermore, the mass of the object is independent of the moduli $\zeta$
-- there is no analog of the shift of the Hamiltonian found in \egs.
(The analogous $|\zeta|^2$ term appears in the combination $|z^1z^2-\zeta|^2$.)

All this seems to add up to consistent space-time physics for the
fractional brane.
Consider $0$-branes in \IIa\ for definiteness -- the fractional $0$-brane
has half the charge of the D$0$-brane and its mass is determined by the
BPS bound for this charge, independent of the moduli.

Analogy with the interpretation for conventional orbifolds
suggests that these objects are $p+2$-branes wrapped about a hidden two-cycle.
Here the complete resolution left a conifold singularity, and if we
excise this singularity we have a space with homology two and three-cycles.
On the other hand there is no reason why excising the singularity is the
right geometrical interpretation and as we mentioned at the beginning
of this section the closed string spectrum has no corresponding states;
furthermore a brane wrapped about a two-cycle would come with a $\BZ$
conserved winding number, which this does not.  The conclusion is that
a purely geometrical picture is inadequate.

\newsec{Conclusions}

We showed how to formulate D-brane world-volume theories on orbifolds
with discrete torsion and worked out the simplest non-trivial example
of $\BC^3/\BZ_2\times \BZ_2$, gaining a good space-time description of the 
orbifold and its resolution which agrees with that suggested by Vafa
and Witten.  The moduli in this case are complex structure moduli which
enter into the superpotential of the D-brane theory.  Turning them on
partially resolves the singularity leaving a conifold singularity.

We also discovered a new BPS ``fractional brane'' which is
bound to the singularity.  It has some similarity with the fractional
branes of conventional orbifolds but does not have a clear interpretation
as a brane wrapped around a hidden cycle.
Applying the mirror symmetry discussed in \vw, we
also predict the existence of fractional $3$-branes in the \IIb\ string
on the $T^6/\BZ_2\times \BZ_2$ orbifold without discrete torsion.

Other examples can be worked out
\inprog; for example $\BZ_n\times \BZ_n$ with the elementary cocycle
is described by a $U(n)$ gauge theory.

Another lesson we can draw from this is that discrete torsion is not
just an attribute of string theory but also exists in M theory.
For example, taking the eleven-dimensional limit to be described
by the D$0$-branes, the difference between the two compactifications
does not disappear in the large $R_{11}$ limit.  On the other hand
the structure of the argument in section 2 suggests that it disappears
in the large $N$ limit and hence would only be visible quantum mechanically
in M theory, and also in an application such as \ks.

In the ``first superstring revolution,'' it was realized that target
spaces for string theory need not have any geometric interpretation --
any conformal field theory could be used to define the string world-sheet.
However, the trend in the ``second superstring revolution'' has been to
find geometric (especially, algebraic geometric) pictures for
everything.

As is discussed in \vw, discrete torsion has not been given
a satisfactory geometric picture.  The results here suggest that
noncommutative geometry is a more relevant framework.
As a concrete goal in this direction, we propose the problem of formulating
a definition of noncommutative space which admits as a
distinct example each space with a choice of discrete torsion which
can appear in string theory.

\medskip

We would like to thank Ken Intriligator and the UCSD Physics
Department for their hospitality.

This work was supported in part by DOE grant DE-FG02-96ER40559.

\listrefs
\end